\RequirePackage[l2tabu,orthodox]{nag}
\RequirePackage{fixltx2e}
\documentclass[reprint, amsmath, amssymb, superscriptaddress]{revtex4-1}

\usepackage[colorlinks]{hyperref}
\usepackage[all]{hypcap}
\usepackage{url}
\usepackage{graphicx}
\usepackage[dvipsnames, table]{xcolor}
\usepackage{microtype}
\usepackage[capitalise]{cleveref}
\usepackage{natbib}
\usepackage{float}
\usepackage{tabularx}
\usepackage{array}
\usepackage{booktabs}
\usepackage{pgffor}
\newcolumntype{Y}{>{\centering\arraybackslash}X}

\makeatletter

    \def\CT@@do@color{%
      \global\let\CT@do@color\relax
            \@tempdima\wd\z@
            \advance\@tempdima\@tempdimb
            \advance\@tempdima\@tempdimc
    \advance\@tempdimb\tabcolsep
    \advance\@tempdimc\tabcolsep
    \advance\@tempdima2\tabcolsep
            \kern-\@tempdimb
            \leaders\vrule
                    \hskip\@tempdima\@plus  1fill
            \kern-\@tempdimc
            \hskip-\wd\z@ \@plus -1fill }
    \makeatother

\makeatletter
\newcommand*{\balancecolsandclearpage}{%
  \close@column@grid
  \clearpage
  \twocolumngrid
}
\makeatother

\begin{document}

\title{Theory of chirped photonic crystals in biological broadband reflectors}
\author{Caleb Q. Cook}\email{calebqcook@gmail.com}
\affiliation{Department of Physics, Harvard University, Cambridge, Massachusetts 02138, USA}
\affiliation{Perimeter Institute for Theoretical Physics, Waterloo, Ontario N2J 2Y5, Canada}
\author{Ariel Amir}\email{arielamir@seas.harvard.edu}
\affiliation{School of Engineering and Applied Sciences, Harvard University, Cambridge, Massachusetts 02138, USA}
\date{\today}

\begin{abstract}
One-dimensional photonic crystals with slowly varying, i.e. `chirped', lattice period are responsible for broadband light reflectance in many diverse biological contexts, ranging from the shiny coatings of various beetles to the eyes of certain butterflies.  We present a quantum scattering analogy for light reflection from these adiabatically chirped photonic crystals (ACPCs) and apply a WKB-type approximation to obtain a closed-form expression for the reflectance. From this expression we infer several design principles, including a differential equation for the chirp pattern required to elicit a given reflectance spectrum and the minimal number of bilayers required to exceed a desired reflectance threshold. Comparison of the number of bilayers found in ACPCs throughout nature and our predicted minimal required number also gives a quantitative measure of the optimality of chirped biological reflectors. Together these results elucidate the design principles of chirped reflectors in nature and their possible application to future optical technologies.

\end{abstract}

\maketitle

Many of the brilliant colors throughout nature are produced via the interaction of light and microscopic structures, a phenomenon known as structural coloration \cite{OpticsOfLife, KinoshitaBook, Kinoshita2008}. In particular, certain species of beetles, butterflies, birds, spiders, and fish take advantage of structural color phenomena to achieve their striking appearance \cite{Land1972, Vukusic2003, Smith2009, Biro2011}. In many of these cases color arises even in the absence of pigments through the layered application of materials with differing refractive indices, structurally arranged so as to yield the desired color via light interference.
 
Perhaps the simplest such structure is a one-dimensional photonic crystal, i.e. a stack of dielectric slabs with alternating high and low refractive indices whose periodicity leads to a band structure analogous to the electronic band structure of semiconductors \cite{Photonic}. The implication of this analogy is the existence of photonic stopbands in which incident light of certain wavelengths is highly reflected due to its inability to propagate within the crystal \cite{Amir2013}. By preferentially reflecting a narrow range of wavelengths, photonic crystals can reliably produce a characteristic color when illuminated with white light.

\begin{figure}[t!]
	\centering
	\includegraphics[width=\columnwidth]{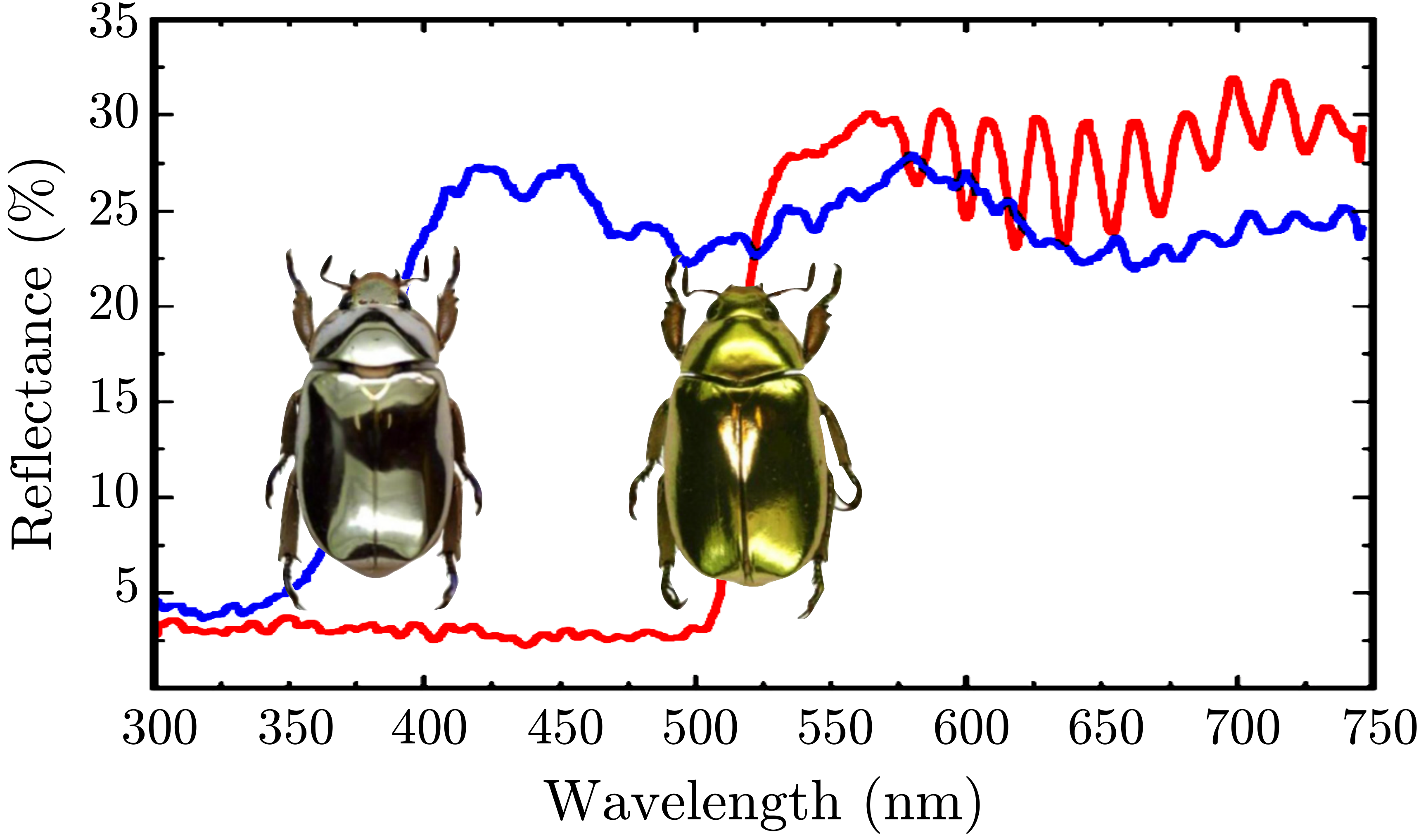}
	\caption{
	\textit{C. limbata} (left) and \textit{C. aurigans} (right) displaying brilliant  silver and gold color, respectively, due to reflection from ACPCs in their elytra. Figure adapted from Ref. \cite{Fernandez2011}.
	}
	\label{fig:beetles}
\end{figure}

\begin{figure}[t!]
	\centering
	\includegraphics[width=\columnwidth]{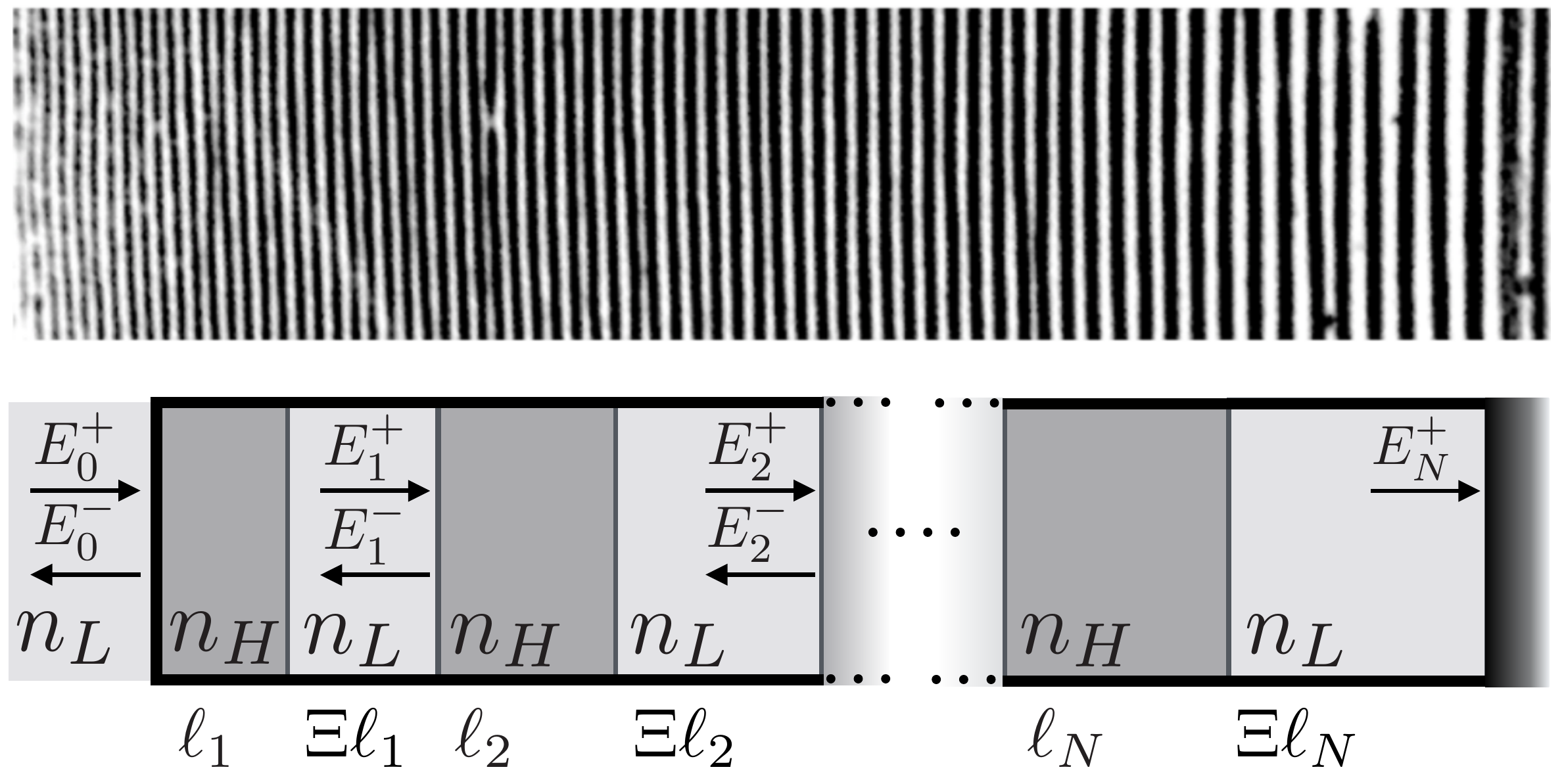}
	\caption{
	\textit{Top}: TEM image of an ACPC in the eye of the crustacean \textit{C. refulgens}. Stained regions show layers, each on the order of 100 nm thick, of differing refractive index. Image taken from Ref. \cite{Nishida2002}. 
	\textit{Bottom}: Our model ACPC. Adiabatic chirp requires that the sequence of optical thicknesses $\ell_m$ varies slowly. The forward- and backward-propagating fields $E^\pm_m$ are also shown, with an absorbing layer at the back enforcing $E_N^-=0$.
	\label{fig:chirps}
	}
\end{figure}

A more sophisticated approach involves gradually changing the period of the crystal. These "adiabatically chirped" photonic crystals (ACPCs) act as a locally periodic crystal for a range of wavelengths and can therefore reflect a broad swath of colors strongly, allowing for reflection spectra more complex than monochromatic color, e.g. the shiny shells of certain beetles \cite{Neville1977, Parker1998, Azo2015} and eye glow of some butterflies \cite{Miller1968, Ribi1981,Labhart2009}. The striking colors that can arise from ACPCs are shown in the case of gold and silver beetles in \cref{fig:beetles}. An ACPC in the eye of a deep sea crustacean is shown in \cref{fig:chirps}, where the chirped structure is signaled by gradual variations in the layer thickness.

Replication of natural photonic structures has already yielded marketable benefit in the manufacture of paints, fabrics, displays, and other technologies \cite{Vukusic2003, Yu2013}. ACPCs in particular are well-suited for applications requiring complex, broadband reflection spectra, such as light manipulation in white LEDs \cite{Yu2013}, dispersion control of ultrashort laser pulses \cite{Matuschek1998}, and tailored optical behavior in fiber gratings \cite{Erdogan1997}. Better understanding the design of biological photonic structures such as ACPCs may therefore provide inspiration for future technologies.

There are many well-known tools for calculating the reflectance of dielectric multilayers, including transfer matrix methods and recursion techniques \cite{Abeles1950,Amir2013,KinoshitaBook}. Although numerically exact, these methods are iterative, and when there are many layers involved, as in most biological applications, carrying out these computations does not solve the inverse problem of designing reflectors with tailored optical behavior. In this Letter, we instead derive an approximate, but closed-form, expression for the reflectance of ACPCs by matching Bloch wave expressions for the electric field across stopband regions via a \emph{discrete} Wentzel-Kramers-Brillouin (WKB) method. Analysis of this closed-form expression gives insight into several of the underlying the design principles of ACPCs.

Matuschek \textit{et. al.} \cite{Matuschek1998} apply a related WKB approximation to study ACPCs for ultrafast laser application. However, their work makes several simplifying assumptions, such as ignoring reflections from the back of the multilayer and thereby assuming near perfect reflection. By instead performing asymptotic matching in full, we obtain a reflectance expression that gives better agreement with numerically exact results. Our method also follows a more elementary approach and does not invoke a needlessly cumbersome coupled-mode formalism \cite{Matuschek1997}.

Consider a stack of $2N$ dielectric slabs of refractive indices $n_L < n_H$ arranged in an alternating fashion. We assume that within the $m$-th $\{n_L,\,n_H\}$ bilayer, the slabs have fixed optical thickness ratio $\Xi=\left(\ell_{L}\right)_{m}/\left(\ell_{H}\right)_{m}$. The sequence of optical thicknesses $\ell_{m}\equiv\left(\ell_{H}\right)_{m}=\left(\ell_{L}\right)_{m}/\Xi
$ then encodes the chirp of the multilayer, which we assume to be monotonic and slowly-varying (compared to the wavelength of light) with bilayer index $m$. We also assume zero back-reflectance at the end of the ACPC, which models a final inhomogeneous absorptive layer, common to biological multilayers, whose light-diffusing character serves to strongly limit the amount of coherent radiation directed back into the structure \cite{Parker1998,Azo2015}. A depiction of the model is shown in \cref{fig:chirps}.

In the transfer matrix method \cite{Abeles1950}, one decomposes the electric field at each bilayer interface $m$ into forward- and backward-propagating components $\mathbf{E}_m=\left[\begin{smallmatrix}E_m^{+}\\E_m^{-}\end{smallmatrix}\right]$. The fields $\mathbf{E}_m$ at bilayer interfaces are related to one another via a transfer matrix product
\begin{equation}
	\mathbf{E}_{m} = F_m\left(\lambda\right)\mathbf{E}_{m+1}
	\label{eq:vecRecm}.
\end{equation}
For incident light of wavelength $\lambda$, the $m$-th bilayer transfer matrix $F_m(\lambda)=F(x)$ of our model is \cite{EWA} 
\begin{widetext}
\begin{equation}
\label{eq:transferMatrixm}
F\left(x\right)=\frac{1}{1-\rho^{2}}\left[\begin{array}{cc}
e^{i\pi x}-\rho^{2}e^{-i\pi\left(\frac{1-\Xi}{1+\Xi}\right)x} & -2i\rho e^{-i\pi\left(\frac{\Xi}{1+\Xi}\right)x}\sin\left(\frac{\pi x}{1+\Xi}\right)\\
2i\rho e^{i\pi\left(\frac{\Xi}{1+\Xi}\right)x}\sin\left(\frac{\pi x}{1+\Xi}\right) & e^{-i\pi x}-\rho^{2}e^{i\pi\left(\frac{1-\Xi}{1+\Xi}\right)x}
\end{array}\right]
\end{equation}
\end{widetext} 
where  $x=x_m(\lambda)=\left(1+\Xi\right)\ell_m/(\lambda/2)$, and $\rho=(n_H-n_L)/(n_H+n_L)$ is the Fresnel reflection coefficient, which we assume to be small as is the case in nearly all biological ACPCs (see \cref{table:scalingComp}).

When $x$ is approximately integer, the transfer matrix recursion \labelcref{eq:vecRecm} has evanescent electric field solutions, corresponding to high reflectance of incident light. Wavelengths $\lambda$ and bilayers $m$ for which $x_m\left(\lambda\right)$ is approximately integer therefore define the \emph{stopband} of the ACPC. Conversely, when $x$ lies between integers \cref{eq:vecRecm} has oscillatory solutions, corresponding to low reflectance and hence defining the \emph{passbands}. The bilayer indices $m_{t}(\lambda)$ which separate stopband/passband regions define the \emph{turning points}. To avoid reflections between multiple stopbands and simplify our analysis, we assume just a single stopband $m\in\left[m_1\left(\lambda\right),m_2\left(\lambda\right)\right]$ for all considered wavelengths $\lambda$. We also assume an $x\approx 1$ stopband, since these require minimal layer thickness and are therefore most prevalent in biological systems.  

In the case of no chirp $\ell_m=\ell$, the vector recursion \labelcref{eq:vecRecm} has exact, Bloch wave solutions $\mathbf{E}_{m}^{\pm}=\exp\left[-m\ln\mu^{\pm}\right]\mathbf{w}^{\pm}$ where $\mathbf{w}^\pm_m=\mathbf{w}^\pm$ are the transfer matrix eigenvectors with corresponding eigenvalues $\mu^\pm_m=\mu^\pm$. A natural generalization of these solutions are the modulated Bloch (mBloch) waves $\mathbf{E}_m^\pm=\exp[{-\int^m\ln\mu^\pm_k\,\mathrm{d}k}]\,\mathbf{w}^\pm_m$, which approximately solve \cref{eq:vecRecm} for adiabatic chirp. Introducing the local Bloch phase 
$\varphi_m=\cos^{-1}[\text{Tr}(F_m)/2]$, approximate solutions to \cref{eq:vecRecm} are then given by linear combinations of mBloch waves
\begin{equation}
	\mathbf{E}_{m}=\mathcal{C}^{+}\exp\left[i\int_{m}^{m_{t}}\varphi_{k}\,\mathrm{d}k\right]\mathbf{w}_{m}^{+}+\mathcal{C}^{-}\exp\left[i\int_{m_{t}}^{m}\varphi_{k}\,\mathrm{d}k\right]\mathbf{w}_{m}^{-}
	\label{eq:blochm}
\end{equation}
with constant coefficients $\mathcal{C}^{\pm}\in\left\{ \left\{ A,B\right\} ,\left\{ C,D\right\} ,\left\{ F,G\right\} \right\}$ in the three regions of the ACPC (passband$\rightarrow$ stopband$\rightarrow$ passband) separated by the turning points $m_{t=1,2}$.

\begin{figure}[t!]
	\centering
	\includegraphics[width=\columnwidth]{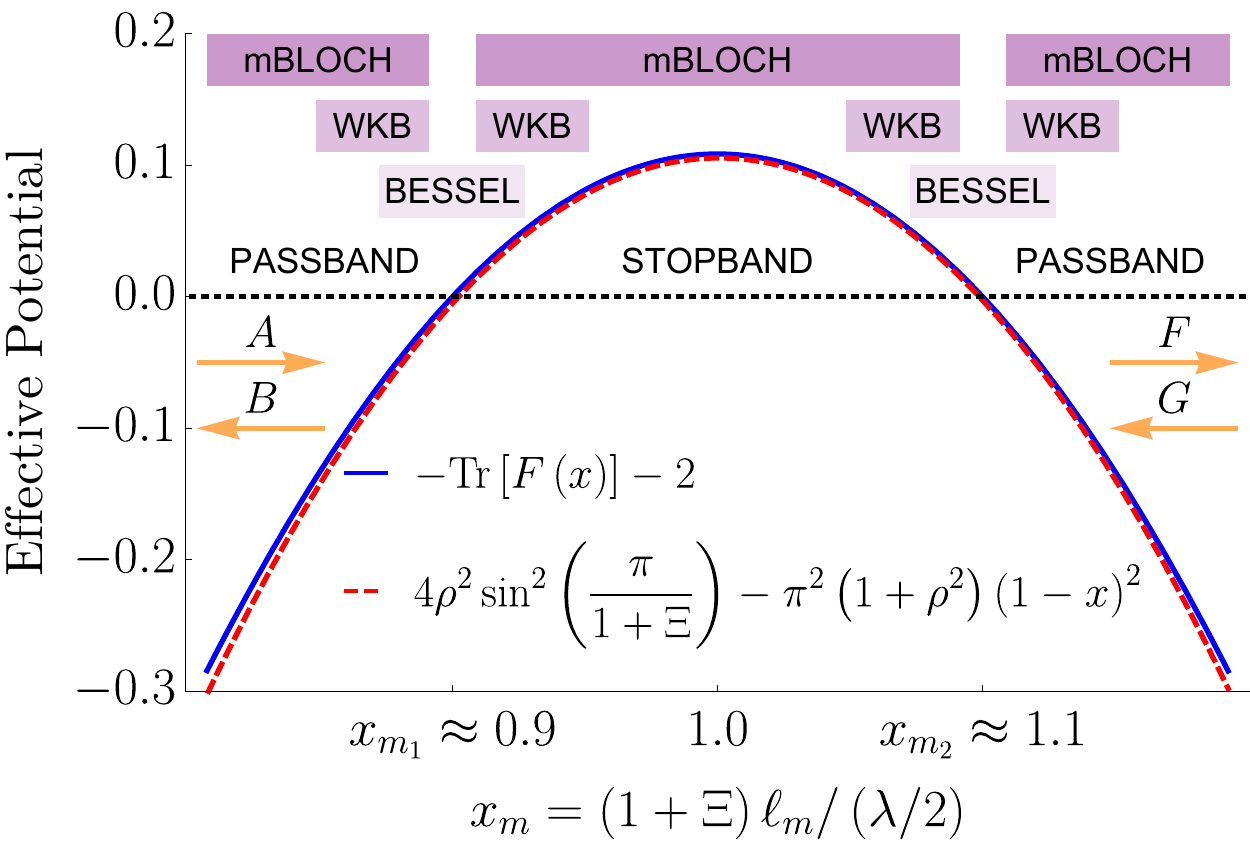}
	\caption{Plot of the stopband potential $-\text{Tr}[F(x)]-2$ and its quadratic expansion about $x=\Xi=1$ and $\rho=0$ in the quantum scattering analogy (detailed in \cref{sec:tight} of the Supplementary Material) for an ACPC with $n_L=1.0$, $n_H=1.4$, and $\Xi=0.75$. The regions of validity for various approximate expressions for $E^{\pm}_m$ are sketched. The WKB matching formulae (derived in \cref{sec:WKB} of the Supplementary Material) relate the pre-stopband coefficients $A$, $B$ and the post-stopband coefficients $F$, $G$ in the mBloch wave solution \labelcref{eq:blochm}.}
	\label{fig:patching}
\end{figure}

Higher order corrections to the mBloch wave solution \labelcref{eq:blochm} are obtained in \cref{sec:reflection} of the Supplementary Material by demanding local energy conservation in the multilayer. These corrections diverge at turning points $m_{1,2}$, which we overcome by studying \cref{eq:vecRecm} in more detail near these points. In particular, adiabatic chirp implies that near a given bilayer $m$, the fields $E^\pm_m$ approximately satisfy the recurrence
\begin{equation}
	\label{eq:recRelationm}
	E^{\pm}_{m+1}\approx \text{Tr}(F_m)E^{\pm}_m-E^{\pm}_{m-1}
\end{equation}
which is formally a discrete Schr{\"o}dinger equation in a potential determined by $\text{Tr}(F_m)$ (see \cref{sec:tight} of the Supplementary Material for details). We therefore apply a discrete WKB method \cite{Dingle1968, Braun1993} to approximately solve \cref{eq:recRelationm} near the turning points $m_{1,2}$ and match the mBloch wave \labelcref{eq:blochm} coefficients $\mathcal{C}^\pm$ across the ACPC stopband. We sketch this procedure in \cref{fig:patching}.

\begin{figure}[t!]
	\centering
	\includegraphics[width=\columnwidth]{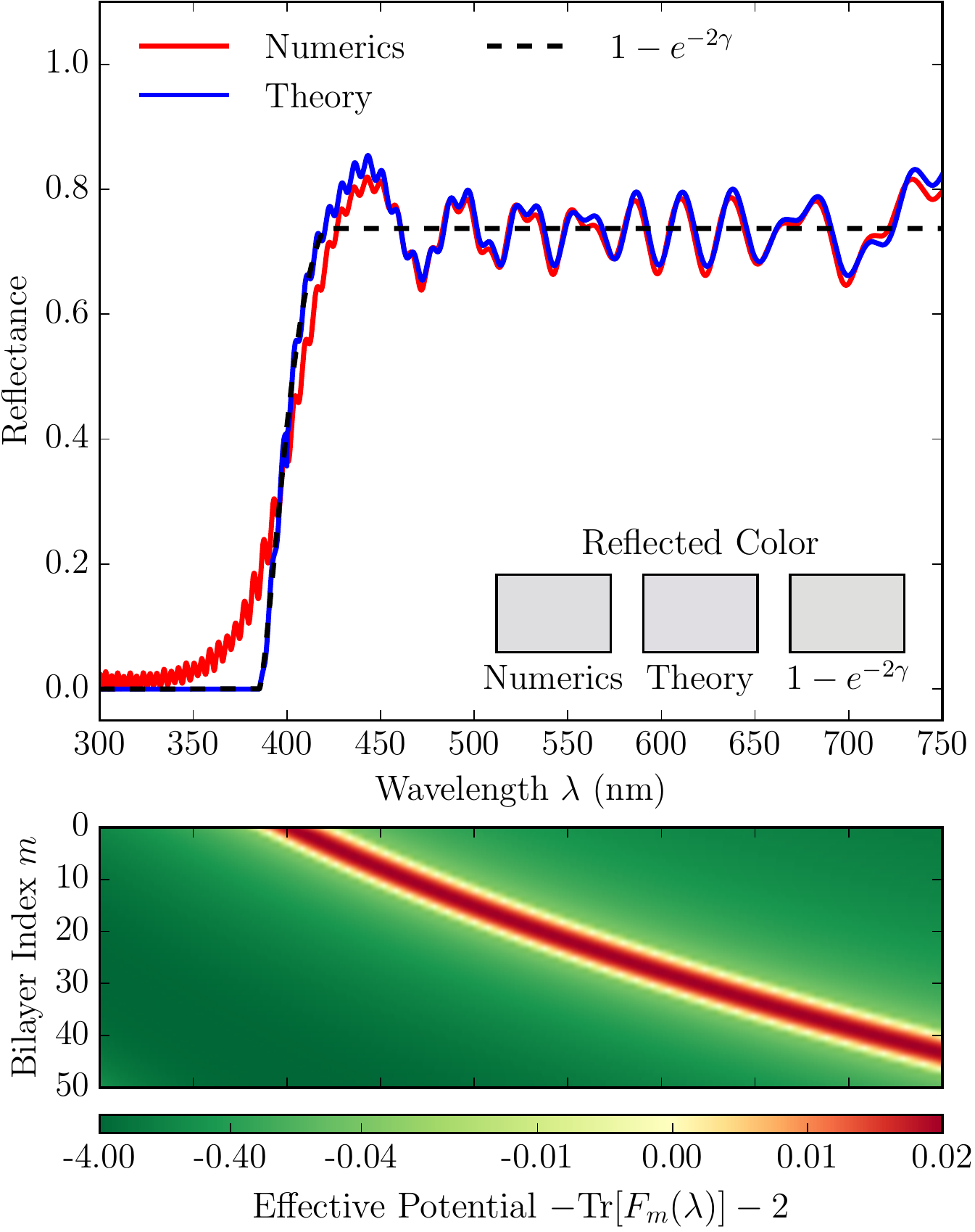}
	\caption{
	An exponentially chirped photonic crystal with $n_L=1.34$, $n_H=1.54$, and $\Xi=1$. \textit{Top}: A comparison between the numerically exact transfer matrix (red), first-order WKB (\cref{eq:refl}, blue), and zeroth-order WKB (Eq. (\ref{eq:Rtildem}), dashed) reflection spectra is shown.
	\textit{Bottom}: The analagous quantum potential of the ACPC is plotted for each incident wavelength.
	}
	\label{fig:reflComp}
\end{figure}

The ratio of fields $E^-_0/E^+_0$ in the mBloch solution \labelcref{eq:blochm} is fixed by enforcing the absorber condition $E_N^-=0$ and applying the discrete WKB matching formulae, which are characterized by the tunneling exponent
\begin{align}
\begin{split}
\label{eq:gammam}
	\gamma(\lambda) & =\int_{m_{1}(\lambda)}^{m_{2}(\lambda)}\cosh^{-1}\left\{-\dfrac{\text{Tr}[F_m(\lambda)]}{2}\right\}\,\mathrm{d}m\\
 	& \approx \frac{\lambda}{q\left(\lambda \right)}\sin^{2}\left(\frac{\pi}{1+\Xi}\right)		\rho^{2}+\mathcal{O}\left(\rho^{3}\right)
\end{split}
\end{align}
where $q\left(\lambda \right)\equiv\left(1+\Xi\right)\dfrac{\mathrm{d}\ell}{\mathrm{d}m}\left(\ell=\frac{\lambda/2}{1+\Xi}\right)
$ is the chirp rate at the center of stopband. In approximating \cref{eq:gammam}, we have used the quadratic expansion of $\text{Tr}(F_m)$ shown in \cref{fig:patching}. The reflectance $R=|E^-_0/E^+_0|^2$ of ACPCs calculated via this discrete WKB matching procedure is then given to lowest order in $\rho$ by
\begin{equation}
	\label{eq:Rtildem}
	\widetilde{R}\left(\lambda\right)=1-\exp\left[-2\gamma(\lambda)\right],
\end{equation}
analogous to the form found in quantum scattering \cite{MerzbacherQM}. This calculation is detailed in \cref{sec:reflection} of the Supplementary Material, where the next-order WKB reflectance is also given in \cref{eq:refl}. A comparison between the numerically exact reflection spectrum of an ACPC and our corresponding WKB results \labelcref{eq:Rtildem,eq:refl} is shown in \cref{fig:reflComp}. As \cref{fig:reflComp} demonstrates, the WKB expression $\widetilde{R}(\lambda)$ \labelcref{eq:Rtildem} in fact captures a moving average of the exact reflectance spectrum. 

Dimensional analysis shows that $q(\lambda)$ in \cref{eq:gammam}, when written in terms of the wavelengths $\lambda_{\text{min}}=2\left(1+\Xi\right)\ell_{0}$ and $\lambda_{\text{max}}=2\left(1+\Xi\right)\ell_{N}$ over which an ACPC reflects strongly, is independent of the optical thickness ratio $\Xi$. The tunneling exponent $\gamma(\lambda)$ \labelcref{eq:gammam} and hence the average reflectance $\widetilde{R}(\lambda)$ \labelcref{eq:Rtildem} is therefore maximal for $\Xi=1$, i.e. when the optical thicknesses are equal in each bilayer. For periodic $\ell_m=\ell$ reflectors, the above fact is well-known, and such multilayers have been dubbed "ideal" \cite{Land1972}. Our results therefore generalize the claim that ideal bilayers elicit maximal reflectance to the case of periodicity-breaking adiabatic chirp.

\begin{table*}[t!]

\begin{tabularx}{\textwidth}{l l l Y l l >{\boldmath}Y >{\boldmath}l X >{\boldmath}l X l}
\toprule

Adiabatically Chirped Photonic Crystal & $n_{L}$ & $n_{H}$ & $\Xi$ & $\lambda_{\text{min}}$ & $\lambda_{\text{max}}$ & \multicolumn{1}{c}{$N$} & \multicolumn{2}{Y}{$N_{\text{exp}}^{\gamma\sim1}$} & \multicolumn{2}{Y}{$N_{\text{lin}}^{\gamma\sim1}$} & Ref.\tabularnewline

\midrule

\emph{Aspidomorpha tecta} (beetle) elytra & $1.40$ & $1.68$ & $0.79$ & $450$ & $850$ & $\sim44$  & $40$ & $\left(-9\%\right)$ & $56$ & $\left(+27\%\right)$ & \cite{Parker1998} \tabularnewline

\emph{Chrysina aurigans} (beetle) elytra & $1.57$ & $1.80$ & $1.10$ & $520$ & $1000$ & $\sim94$ & $71$ & $\left(-24\%\right)$ & $100$ & $\left(+6\%\right)$ & \cite{Libby2014} \tabularnewline

\emph{Cephalophanes refulgens} (crustacean) eyes & $1.34$ & $1.54$ & $0.87$ & $350$ & $700$ & $\sim75$ & $73$ & $\left(-3\%\right)$ & $105$ & $\left(+40\%\right)$ & \cite{Nishida2002} \tabularnewline

\emph{Danaella} sp. (crustacean) antennae & $1.34$ & $1.54$ & $0.86$ & $370$ & $700$ & $\sim74$ & $67$ & $\left(-9\%\right)$ & $94$ & $\left(+27\%\right)$ & \cite{Parker1999} \tabularnewline

\rowcolor{yellow!20}
\emph{Danaus plexippus} (butterfly) tapeta & $1.00$ & $1.40$ & $0.63$ & $300$ & $550$ & $\sim44$ & $12$ & $\left(-73\%\right)$ & $17$ & $\left(-61\%\right)$ & \cite{Labhart2009} \tabularnewline

\rowcolor{yellow!20}
\emph{Vanessa cardui }(butterfly) tapeta & $1.00$ & $1.50$ & $0.64$ & $320$ & $680$ & $\sim35$ & $11$ & $\left(-63\%\right)$ & $16$ & $\left(-54\%\right)$ & \cite{Briscoe2003} \tabularnewline

\bottomrule
\end{tabularx}

\caption{Comparison between the number of bilayers $N$ found in several biological ACPCs and our predicted minimal number $N^{\gamma\sim1}_\text{exp/lin}$ \labelcref{eq:Nminm}. The percent difference between $N$ and $N^{\gamma\sim1}_\text{exp/lin}$ is given in the relevant column. All wavelengths $\lambda_\text{min, max}$ given in nanometers. The optical thickness ratio $\Xi$ in each case is estimated via analysis of TEM images given in the corresponding reference (see \cref{sec:imaging} of the Supplementary Material).} 
	
\label{table:scalingComp}
\end{table*}

\begin{figure}[t!]
	\centering
	\includegraphics[width=\columnwidth]{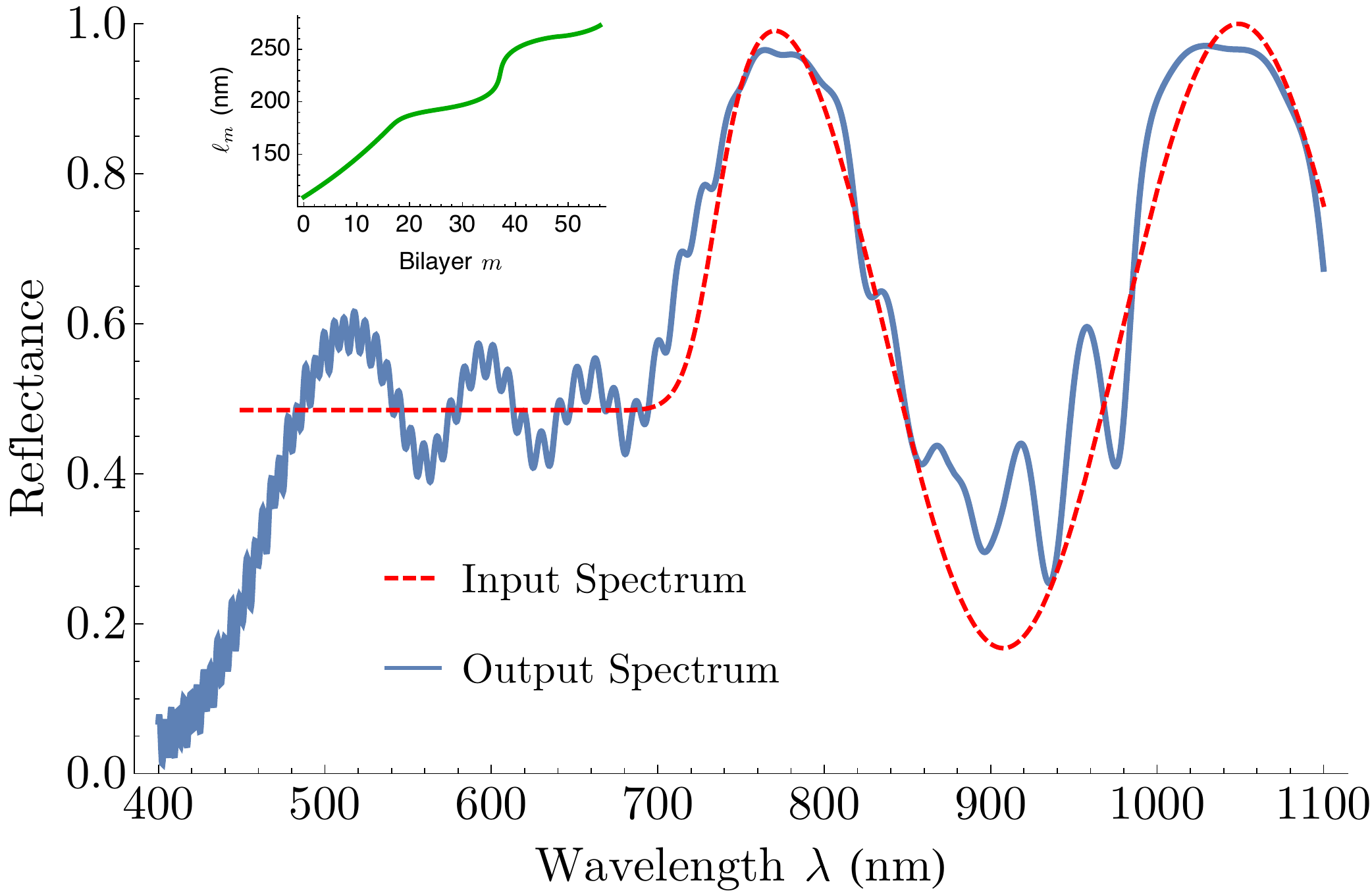}
	\caption{
	An ACPC with a designed average reflectance spectrum. The desired spectrum (red) is put into the chirp differential equation \labelcref{eq:designDEm} with $n_H=1.54$ and $n_L=1.34$, which is solved numerically to give the chirp function $\ell_m$ plotted in the inset (green). The resulting exact reflectance (blue) of the computed chirp $\ell_m$ is then found numerically via transfer matrices and plotted against the input spectrum.
	}
	\label{fig:designed}
\end{figure}

Our results can also be used to solve the following design problem: given a desired average reflectance spectrum $\widetilde{R}(\lambda)$, what chirp function $\ell_m$ is required to elicit $\widetilde{R}(\lambda)$? Solving \cref{eq:gammam,eq:Rtildem} for $\mathrm{d}\ell/\mathrm{d}m$ reveals that this question can be answered by simply solving the differential equation
\begin{equation}
	\label{eq:designDEm}
	\frac{\mathrm{d}\ell}{\mathrm{d}m}=\frac{-4\rho^{2}\sin^{2}\left(\frac{\pi}{1+\Xi}\right)}{\log\left[1-\widetilde{R}\Big(\lambda=2\left(1+\Xi\right)\ell\Big)\right]}\ell.
\end{equation}
The chirp differential equation \labelcref{eq:designDEm} offers a novel method of designing multilayer reflectors of tailored optical behavior, and we exhibit such a design spectrum in \labelcref{fig:designed}. Moreover, \cref{eq:designDEm} shows that achieving a flat spectrum $\widetilde{R}(\lambda)=\widetilde{R}$  requires exponential chirp of the form $\ell_m = \ell_0 \left(\ell_N/\ell_0\right)^{m/N}$. Constant reflectance and hence exponential chirp are of special interest since reflectance above some threshold value is minimally achieved with a flat spectrum.

By computing $q(\lambda)$ in \cref{eq:gammam} for specified chirp functions, we can in fact calculate the minimal number of bilayers $N$ required to exceed some reflectance $\widetilde{R}$ over a given range of wavelengths. For comparison, we carry out this calculation for both exponential $\ell_m = \ell_0 \left(\ell_N/\ell_0\right)^{m/N}$ and linear $\ell_m=\ell_0+(\ell_N-\ell_0)(m/N)$ chirps, which gives the minimal bilayer numbers
\begin{equation}
N=\frac{\log\left(\frac{1}{1-\widetilde{R}}\right)}{4\rho^{2}\sin^{2}\left(\frac{\pi}{1+\Xi}\right)}\times\begin{cases}
\log\left(\lambda_{\text{max}}/\lambda_{\text{min}}\right), & \left(\text{exponential}\right)\\
\lambda_{\text{max}}/\lambda_{\text{min}}-1, & \left(\text{linear}\right)
\end{cases}
\label{eq:Nminm}
\end{equation}
Note that (i) \cref{eq:Nminm} is minimized in either case for $\Xi=1$ and (ii) $N_\text{exp}\leq N_\text{lin}$, consistent with our earlier conclusions. We also emphasize that the design principles contained in \cref{eq:designDEm,eq:Nminm} are not inferable via traditional, numerical (e.g. transfer matrix) approaches.

Comparison between the number of bilayers $N$ found in several biological ACPCs and our predicted minimal number $N_\text{exp/lin}$ \labelcref{eq:Nminm} gives a quantitative measure of the optimality of ACPCs in nature. In order to make a concrete comparison, we fix an arbitrary high reflectance benchmark of $\gamma\sim1$, i.e. $\widetilde{R}\sim 86\%$. This comparison is shown in \cref{table:scalingComp} for several ACPCs in nature and reveals that biological ACPCs are generally less optimal than exponentially chirped reflectors but generally more optimal than their linearly chirped counterparts.

Exceptions to the above conclusion are the butterfly tapeta highlighted in \cref{table:scalingComp}, which have many more bilayers than expected per our analysis. This is consistent, however, with the fact that optical components such as tapeta likely require near perfect reflection, corresponding to $\gamma>1$ and a greater number of bilayers in \cref{eq:Nminm}. Indeed, this explanation is supported by the fact that the reported reflection spectra of \emph{D. plexippus} \cite{Labhart2009} and \emph{V. cardui} \cite{Briscoe2003} tapeta are essentially unity $\widetilde{R}\approx1$ over the range of wavelengths $[\lambda_\text{min}, \lambda_\text{max}]$.

In summary, our closed-form results allow for tailoring ACPCs of desired reflectance spectra and estimating the optimality of chirped reflectors in nature. More generally, the mathematical methods developed herein are applicable to coupled wave propagation in slowly-varying media and therefore also relevant to the design of mechanical and acoustic metamaterials.

We thank Dvir Gur, Pete Vukusic, Gary Bernard, and Mark Arildsen for helpful
discussions. C.Q.C. acknowledges support from the Harvard College Research Program and the Perimeter Institute. Research at PI is supported by the Government of Canada through Industry Canada and by the Province of Ontario through the Ministry of Research and Innovation. A.A. thanks the support of the Alfred P. Sloan Foundation.

% Bibliography
\bibliography{CookAmir2016}
\balancecolsandclearpage

%%%%%%%%%% Merge with supplemental materials %%%%%%%%%%
\pagebreak
\onecolumngrid
\begin{center}
\textbf{\large Theory of chirped photonic crystals in biological broadband reflectors:}
\\ 
\textbf{\large \textit{Supplementary Material}}
\end{center}
\twocolumngrid
%%%%%%%%%% Merge with supplemental materials %%%%%%%%%%
%%%%%%%%%% Prefix a "S" to all equations, figures, tables and reset the counter %%%%%%%%%%
\setcounter{equation}{0}
\setcounter{figure}{0}
\setcounter{table}{0}
\setcounter{page}{1}
\makeatletter
\renewcommand{\theequation}{S\arabic{equation}}
\renewcommand{\thefigure}{S\arabic{figure}}
%%%%%%%%%% Prefix a "S" to all equations, figures, tables and reset the counter %%%%%%%%%%

\section{DERIVATION OF WKB REFLECTANCE}
\label{sec:reflection}

In this section, we outline the calculation of the WKB expression for the reflectance of adiabatic chirped photonic crystals presented in the main text. We also give the WKB reflectance calculated to the next order in $\rho$.

As presented in the main text, the wavelength-dependent bilayer transfer matrix $F_m(\lambda)$, relating the forward- and backward-propagating electric fields 
\begin{equation}
	\mathbf{E}_m=\left[\begin{array}{c}
	E_m^{+}\\
	E_m^{-}
	\end{array}\right]
\end{equation}
via the vector recursion relation
\begin{equation}
	\label{eq:vecRec}
	\mathbf{E}_{m} = F_m(\lambda)\,\mathbf{E}_{m+1},
\end{equation}
for our model takes the form
\begin{equation}
	\label{eq:transferMatrix}
	F_{m}\left(\lambda\right)=\left[\begin{array}{cc}
	X_{m}\left(\lambda\right) & Y_{m}\left(\lambda\right)\\
	Y_{m}^{\star}\left(\lambda\right) & X_{m}^{\star}\left(\lambda\right)
\end{array}\right]
\end{equation}
with complex matrix elements  $X_m(\lambda)$, $Y_m(\lambda)$ satisfying 
$\det F_{m}(\lambda)=\left|X_{m}\right|^{2}-\left|Y_{m}\right|^{2}=1$. The eigenvectors $\mathbf{w}^{\pm}_m$ and corresponding eigenvalues $\mu^{\pm}_m$ of the general transfer matrix $F_m$ are explicitly
\begin{equation}
	\label{eq:eigVecs}
	\mathbf{w}_{m}^{\pm} =  D_m^{\pm}\left[\begin{array}{c}
	1\\
	\left(\dfrac{\mu_{m}^{\pm}-X_{m}}{Y_{m}}\right)
	\end{array}\right]
\end{equation}
and
\begin{equation}
	\label{eq:eigVals}
	\mu_{m}^{\pm} =  \left(\frac{X_{m}+X_{m}^{\star}}{2}\right)\pm\sqrt{\left(\frac{X_{m}+X_{m}^{\star}}{2}\right)^{2}-1},
\end{equation}
where the $D_m^{\pm}$ are eigenvector normalizations, which we will discuss in more detail shortly.

In the main text it is shown that linear combinations of modulated Bloch waves
\begin{equation}
\label{eq:linCombSol}
	\mathbf{E}_{m}=\sum_{i=\pm}\mathcal{C}^{i}\exp\left[-\int^{m}\ln\mu_{k}^{i}\mathrm{d}k\right]\mathbf{w}_{m}^{i}
\end{equation}
approximately solve the transfer matrix recursion \labelcref{eq:vecRec}
in the limit of adiabatic chirp. One could proceed by seeking higher-order corrections to the modulated Bloch wave solution \labelcref{eq:linCombSol}, but we instead invoke energy conservation to find the $m$-dependence of the eigenvector normalizations $D_m^{\pm}$  and thereby extract the dominant effect of these corrections \cite{Garg1998}. From \cref{eq:eigVals,eq:linCombSol} we calculate the Poynting energy flux density $\mathcal{P}_m\propto\left|E_{m}^{+}\right|^{2}-\left|E_{m}^{-}\right|^{2}$ \cite{EWA} to be
\begin{widetext}
\begin{equation}
	\label{eq:poynting}
	\mathcal{P}_{m}\propto\begin{cases}
	\left|\mathcal{C}^{+}\right|^{2}\left|D_{m}^{+}\right|^{2}\left(\dfrac{\mu_{m}^{+}-\mu_{m}^{-}}{\mu_{m}^{+}-X_{m}^{\star}}\right)-\left|\mathcal{C}^{-}\right|^{2}\left|D_{m}^{-}\right|^{2}\left(\dfrac{\mu_{m}^{+}-\mu_{m}^{-}}{\mu_{m}^{-}-X_{m}^{\star}}\right) & \left(\text{passband}\right)\\
	\left(\mathcal{C}^{+}D_{m}^{+}\right)\left(BD_{m}^{-}\right)^{\star}\left(\dfrac{\mu_{m}^{+}-\mu_{m}^{-}}{\mu_{m}^{+}-X_{m}^{\star}}\right)-\left(\mathcal{C}^{+}D_{m}^{+}\right)^{\star}\left(\mathcal{C}^{-}D_{m}^{-}\right)\left(\dfrac{\mu_{m}^{+}-\mu_{m}^{-}}{\mu_{m}^{-}-X_{m}^{\star}}\right) & \left(\text{stopband}\right)
	\end{cases}
\end{equation}
\end{widetext}
which \textit{a priori} defpends on the bilayer index $m$. However, taking the eigenvector normalizations $D^\pm_m$ to scale as 
\begin{equation}
	\label{eq:eigvecNorm}
	D_m^{\pm}\sim\sqrt{\dfrac{\mu_m^{\pm}-X_m^\star}{\mu^+_m - \mu^-_m}}.
\end{equation}
removes the $m$-dependence of $\mathcal{P}_m=\mathcal{P}$ \labelcref{eq:poynting} and therefore enforces constant energy flux density in the multilayer, as required by energy conservation. The modulated Bloch wave solution \labelcref{eq:linCombSol} with the eigenvector normalizations \labelcref{eq:eigvecNorm} is a generalization of the Bremmer method for deriving WKB-type solutions \cite{Bremmer1949, Berry1972}. 

As a turning point $m\to m_{1,2}$ in the multilayer is approached, the transfer matrix eigenvalues \labelcref{eq:eigVals} $\mu_m^+\to\mu_m^-$ converge to one another and the required eigenvector normalizations $D_m^{\pm}$ \labelcref{eq:eigvecNorm} become arbitrarily large. This implies that our assumption of slowly varying transfer matrix eigenvectors $\mathbf{w}^{\pm}_m$ breaks down and our approximate solution \labelcref{eq:linCombSol} no longer holds near the turning points $m\sim m_{1,2}$, even in the case of adiabatic chirp. We must therefore write general solutions of the form \labelcref{eq:linCombSol} separately in the three regions of the multilayer separated by the turning points $m_{1}$ and $m_{2}$, giving

\begin{widetext}
\begin{equation}
	\label{eq:3regionSol}
	\left[\begin{array}{c}
	E_{m}^{+}\\
	E_{m}^{-}
	\end{array}\right]=
	\begin{cases}
	A\exp\left\{ i\int_{m}^{m_{1}}\cos^{-1}\left(\dfrac{a_{k}}{2}\right)\,\mathrm{d}k\right\} \mathbf{w}_{m}^{+}+B\exp\left\{ -i\int_{m}^{m_{1}}\cos^{-1}\left(\dfrac{a_{k}}{2}\right)\,\mathrm{d}k\right\}  \mathbf{w}_{m}^{-} &\left(m<m_{1}\right)\\

	\left(-1\right)^{m}\Bigg[C\exp\left\{ \int_{m_{1}}^{m}\cosh^{-1}\left(-\dfrac{a_{k}}{2}\right)\,\mathrm{d}k\right\}  \mathbf{w}_{m}^{-}+D\exp\left\{ -\int_{m_{1}}^{m}\cosh^{-1}\left(-\dfrac{a_{k}}{2}\right)\,\mathrm{d}k\right\}  \mathbf{w}_{m}^{+}\Bigg] &\left(m_{1}<m<m_{2}\right)\\

	F\exp\left\{ -i\int_{m_{2}}^{m}\cos^{-1}\left(\dfrac{a_{k}}{2}\right)\,\mathrm{d}k\right\}  \mathbf{w}_{m}^{+}+G\exp\left\{ i\int_{m_{2}}^{m}\cos^{-1}\left(\dfrac{a_{k}}{2}\right)\,\mathrm{d}k\right\}  \mathbf{w}_{m}^{-} &\left(m_{2}<m\right)
	\end{cases}
\end{equation}
\end{widetext}
where we have introduced the complex coefficients $A$, $B$, $C$, $D$, $F$, and $G$, in addition to the trace $a_{m}(\lambda)\equiv\text{Tr}\left[F_{m}(\lambda)\right]$ whose magnitude determines the complex structure of the transfer matrix eigenvalues $\mu^{\pm}_m$, so that
\begin{equation}
	\begin{cases}
		\left|a_{m}\left(\lambda\right)\right|<2 \iff m\in\text{passband}\\
		\left|a_{m}\left(\lambda\right)\right|>2 \iff  m\in\text{stopband}\\
		\left|a_{m}\left(\lambda\right)\right|=2 \iff  m\text{ is a turning point}
	\end{cases}
\end{equation}

Since our modulated Bloch wave solution \labelcref{eq:3regionSol} breaks down near turning points $m\sim m_{1,2}$, we return to the exact vector recurrence \labelcref{eq:vecRec}, which may be approximated in the limit of adiabatic chirp as
\begin{equation}
	\mathbf{E}_{m+1}+\mathbf{E}_{m-1}=\left(F_{m}^{-1}+F_{m-1}\right)\mathbf{E}_{m}\approx a_{m}\mathbf{E}_{m}
\end{equation}
in the vicinity of a given bilayer $m$. This implies that we can treat $E^\pm_m$ as decoupled fields each independently satisfying the recurrence relation
\begin{equation}
	\label{eq:recRelation}
	E^{\pm}_{m+1}=a_mE^{\pm}_m-E^{\pm}_{m-1}
\end{equation}
near turning points $m\sim m_{1,2}$. A discrete form of the Wentzel-Kramers-Brillouin (WKB) method \cite{Dingle1968, Braun1978, Wilmott1985, Geronimo1992, Braun1993} then allows us to approximately solve \cref{eq:recRelation} in the vicinity of turning points $m\sim m_{1,2}$ and match the modulated Bloch waves \labelcref{eq:3regionSol} across the stopband region $m\in[m_1, m_2]$. 

The discrete WKB method has traditionally found application in approximate calculations of quantum mechanical energy spectra \cite{Braun1978, Braun1993, Garg1998, Garg1999}. The discrete WKB method in the present context is detailed in \cref{sec:WKB}, with discrete WKB solutions to the difference equation \labelcref{eq:recRelation} given by \cref{eq:WKBeq}. 

Crucially, the discrete WKB solutions \labelcref{eq:WKBeq} share the same functional form as the modulated Bloch waves \labelcref{eq:3regionSol} in the vicinity of a turning point $m\sim m_{1,2}$, since in that limit the transfer matrix eigenvector normalizations \labelcref{eq:eigvecNorm} become
\begin{equation}
	\label{eq:ansatzWKB}
	D_m^+\sim D_m^-\sim(\mu_m^+-\mu_m^-)^{-1/2}=(a_m^2-4)^{-1/4}
\end{equation}
and \cref{eq:3regionSol} subsequently reduces to the discrete WKB solutions \labelcref{eq:WKBeq}. We therefore identify the complex coefficients $A,\ldots,G$ in the modulated Bloch wave solution \labelcref{eq:3regionSol} with the analogous coefficients in the discrete WKB solution \labelcref{eq:WKBeq}. Applying an asymptotic matching procedure to the discrete WKB solution coefficients, and hence the modulated Bloch wave coefficients $A$, $B$, $F$, and $G$, then allows for patching of the modulated Bloch waves \labelcref{eq:3regionSol} across the stopband region.

Asymptotic matching of the discrete WKB solutions is performed in \cref{sec:WKB}, and the resulting connection formulae across an $x_m\approx 1$ (or more generally, $a_m<-2$) stopband $m\in[m_1,m_2]$ are given by
\begin{equation}
	\label{eq:matching}
	\left[\begin{array}{c}
	A\\
	B
	\end{array}\right]=\left[\begin{array}{cc}
	\alpha & \beta\\
	\beta^{\star} & \alpha^{\star}
	\end{array}\right]\left[\begin{array}{c}
	F\\
	G
	\end{array}\right]
\end{equation}
where we have introduced the variables
\begin{align}
	\alpha(\lambda)& = e^{i\left[m_2(\lambda)-m_1(\lambda)\right]\pi}\left[e^{\gamma(\lambda)}+\frac{1}{4}e^{-\gamma(\lambda)}\right]\\
	\beta(\lambda)& = -ie^{-i\left[m_2(\lambda)+m_1(\lambda)\right]\pi}\left[e^{\gamma(\lambda)}-\frac{1}{4}e^{-\gamma(\lambda)}\right]\\
	\label{eq:gamma}
	\gamma(\lambda)& = \int_{m_1(\lambda)}^{m_2(\lambda)}\cosh^{-1}\left[-\dfrac{a_{m}(\lambda)}{2}\right]\,\mathrm{d}m
\end{align}
The parameter $\gamma$ is analogous to the tunneling exponent in the Gamow theory of alpha decay \cite{MerzbacherQM}, in which the probability $T$ of $\text{He}^{2+}$ emission from an atomic nucleus satisfies $T\sim e^{-2\gamma}$. In the present case, $\gamma(\lambda)$ \labelcref{eq:gamma} characterizes how deeply light of wavelength $\lambda$ penetrates an $a_m<-2$ stopband of a chirped photonic crystal.

The terminating absorber condition $E_N^-=0$ assumed in our model gives a linear relation between the coefficients $F$ and $G$ in the  modulated Bloch wave solution \labelcref{eq:3regionSol} and thus a linear relation between the coefficients $A$ and $B$ via the discrete WKB matching formulae \labelcref{eq:matching}. This linear relation fixes $A/B$ and hence the field ratio $r=E^-_0/E^+_0$ through the modulated Bloch wave solution \labelcref{eq:3regionSol}. The reflectance $R=|r|^2$ of ACPCs calculated in this way is given by
\begin{widetext}
\begin{equation}
\label{eq:refl}
	R\left(\lambda\right)=\widetilde{R}\left(\lambda\right)+2\rho\sqrt{\widetilde{R}\left(\lambda\right)}\left[1-\widetilde{R}\left(\lambda\right)\right]\left\{ \dfrac{\sin\left(2\pi\ell_{0}/\lambda\right)}{\sin\left[2\pi\left(1+\Xi\right)\ell_{0}/\lambda\right]}\sin2\phi_{0}-\dfrac{\sin\left(2\pi\ell_{N}/\lambda\right)}{\sin\left[2\pi\left(1+\Xi\right)\ell_{N}/\lambda\right]}\sin2\phi_{N}\right\} +\mathcal{O}\left(\rho^{2}\right)
\end{equation}
\end{widetext}
where we have again introduced the zeroth-order WKB reflectance 
\begin{equation}
\label{eq:Rtilde}
	\widetilde{R}\left(\lambda\right)=1-\exp\left[-2\gamma(\lambda)\right].
\end{equation} 
as well as the phase angles
\begin{align}
	\phi_{0}\left(\lambda\right) & =\int_{0}^{m_{1}\left(\lambda\right)}\cos^{-1}\left[\frac{a_{m}\left(\lambda\right)}{2}\right]\mathrm{d}m-m_{1}\left(\lambda\right)\pi\\
	\phi_{N}\left(\lambda\right) & =\int_{m_{2}\left(\lambda\right)}^{N}\;\,\cos^{-1}\left[\frac{a_{m}\left(\lambda\right)}{2}\right]\mathrm{d}m+m_{2}\left(\lambda\right)\pi
\end{align}

\section{TIGHT-BINDING ANALOGY}
\label{sec:tight}

In this section, we make an analogy between the electric fields $E_{m}^{\pm}$
in an adiabatically chirped photonic crystal and the zero energy
eigenmodes of a one-dimensional chain, described quantum mechanically
by a tight-binding Hamiltonian. Naively applying the continuous WKB method to the quantum chain yields results consistent
with those of the discrete WKB analysis near turning points and offers insight into the functional form of $E_{m}^{\pm}$ and our matching formulae. By taking
a continuum limit of the quantum chain tight-binding Hamiltonian,
we also make concrete the analogy between the reflection of light from
a photonic stopband and quantum scattering from a potential barrier.

We note that the recurrence relation \labelcref{eq:recRelation} is precisely the equation one would obtain by seeking the zero
energy eigenmodes $E_{m}$ of a Hamiltonian
$H$ given by
\begin{equation}
\label{eq:hamiltonian}
	H=\left[\begin{array}{ccccc}
	a_{1} & -1\\
	-1 & a_{2} & \ddots\\
	 & -1 & \ddots & -1\\
	 &  & \ddots & a_{N-1} & -1\\
	 &  &  & -1 & a_{N}
	\end{array}\right].
\end{equation}
We can view the matrix $H$ as a tight-binding Hamiltonian of a finite,
one-dimensional quantum chain with nearest-neighbor hopping elements \cite{AshcroftMermin}.

Let us first begin with the simple case that $a_{m}=a$ is constant.
In this case, we can write the exact eigenmode solution as a discrete
plane wave $E_{m}=e^{\pm i\kappa m}$
with wavevector $\kappa$. Substitution into our recurrence relation \labelcref{eq:recRelation}
then tells us that 
\begin{equation}
	e^{i\kappa}+e^{-i\kappa}=a
\end{equation}
which implies that our wavevector $\kappa$ is given by 
\begin{equation}
	\kappa=\begin{cases}
	\cos^{-1}\left(a/2\right), & a\in\left[-2,2\right]\\
	i\cosh^{-1}\left(a/2\right), & a>2\\
	i\cosh^{-1}\left(a/2\right)+\pi, & a<-2
	\end{cases}.
\end{equation}
The general eigenmode solution $E_{m}$ in the
constant $a_{m}=a$ case is then just a linear superposition
\begin{equation}
	E_{m}=Ae^{i\kappa m}+Be^{-i\kappa m}
\end{equation}
of the discrete plane waves of wavevectors $\pm\kappa$. When $a\in\left[-2,2\right]$
and $\kappa$ is real, the zero energy eigenmode components $E_{m}$
are oscillatory in the bilayer/chain site index $m$. Conversely,
when $\left|a\right|>2$ and $\kappa$ is imaginary, the components
$E_{m}$ are the sum of growing and decaying exponentials
in the index $m$.

In the more general case that $a_{m}$ varies slowly with bilayer/chain
site index $m$, we can write the continuous WKB solution to the above eigenmode
problem as \cite{MerzbacherQM}
\begin{equation}
\label{eq:standardWKB}
	E_{m}\propto\frac{1}{\sqrt{\kappa_{m}}}e^{\pm i\int^{m}\kappa_{k}\mathrm{d}k}.
\end{equation}
Substituting this expression into our recursion relation gives, to
lowest order in the variation of $\kappa_{m}$ with respect to $m$,
the same consistency equation as before 
\begin{equation}
	e^{i\kappa_{m}}+e^{-i\kappa_{m}}=a_{m}
\end{equation}
which implies (variable) wavevector solutions
\begin{equation}
\label{eq:standardWKBtight}
	\kappa_m=\begin{cases}
	\cos^{-1}\left(a_{m}/2\right), & a_{m}\in\left[-2,2\right]\\
	i\cosh^{-1}\left(a_{m}/2\right), & a_{m}>2\\
	i\cosh^{-1}\left(a_{m}/2\right)+\pi, & a_{m}<-2
	\end{cases}.
\end{equation}
Again, the magnitude of $a_{m}$ tells us whether the the eigenmode
components $E_{m}$ are oscillatory or exponential
in the index $m$. 

Comparing the continuum WKB result \labelcref{eq:standardWKB} with the form of $E_{m}$ obtained via the discrete
WKB method \labelcref{eq:WKBeq}, we see that the two differ only by the $m$-dependence
of the passband magnitude $\left|E_{m}\right|$ of the electric
field/zero energy eigenmode. In particular, 
\begin{equation}
	\left|E_{m}\right|=\begin{cases}
	\left[\cos^{-1}\left(a_{m}/2\right)\right]^{-1/2} & \labelcref{eq:standardWKB}\\
	\left(4-a_{m}^{2}\right)^{-1/4} & \labelcref{eq:WKBeq}
	\end{cases}
\end{equation}
in passband regions $\left|a_{m}\right|<2$. However, these two
forms are consistent to lowest order in $\left(a_{m}\pm2\right)$, i.e. near turning points, which is precisely the regime in which the difference equation  \labelcref{eq:recRelation} is applied in \cref{sec:WKB} to match the modulated Bloch waves \labelcref{eq:3regionSol}. This explains the strong similarity (exact agreement) between the standard WKB connection formulae \cite{MerzbacherQM} and the discrete WKB matching formulae across an odd (even) stopband and given in \cref{sec:WKB}.

We can gain further intuition from this tight-binding analogy by rewriting
the tight-binding Hamiltonian $H$ in the continuum limit as
\begin{equation}
	H\to-\frac{\mathrm{d}^{2}}{\mathrm{d}z^{2}}+\left[a\left(z\right)-2\right]=\left(-i\frac{\mathrm{d}}{\mathrm{d}z}\right)^{2}+V\left(z\right)
\end{equation}
where $m\to z$ represents the continuous coordinate parametrizing
the depth in the multilayer and/or length along the quantum chain, and
we have introduced the potential
\begin{equation}
\label{eq:evenPot}
	V\left(z\right)=a\left(z\right)-2.
\end{equation}

Examining the continuum quantum chain potential $V\left(z\right)$ reveals why one obtains evanescent solutions when $a\left(z\right)>2$;
in this case the potential \labelcref{eq:evenPot} is positive, and therefore the zero energy
modes we seek are classically forbidden by energy conservation. In
quantum mechanics, however, such classically forbidden solutions to
$HE\left(z\right)=0$ are allowed, but are exponentially
damped in $z$ due to quantum tunneling \cite{MerzbacherQM}. 

An identical analogy can be made for odd $a\left(z\right)<-2$ stopbands by transforming the recurrence  \labelcref{eq:recRelation} in $E_m$ to the recurrence
\begin{equation}
	-\widetilde{E}_{m+1}+\widetilde{a}_{m}\widetilde{E}_{m}-\widetilde{E}_{m-1}=0
\end{equation}
in the transformed variable $\widetilde{E}_m=(-1)^mE_m$, where have defined $\widetilde{a}(z)=-a(z)$. Repeating the same analysis as before, we arrive at the odd stopband potential
\begin{equation}
	\widetilde{V}\left(z\right)=\widetilde{a}\left(z\right)-2=-a(z)-2
\end{equation}
which is plotted in the main text.

We therefore conclude that stopbands in adiabatically chirped photonic crystals are analogous to the classically forbidden regions on a one-dimensional tight-binding
quantum chain, with light scattering from the former corresponding
to quantum tunneling through the latter.

\onecolumngrid
\section{DERIVATION OF DISCRETE WKB CONNECTION FORMULAE}
\label{sec:WKB}

In this section, we first derive the discrete WKB connection formulae across an even, i.e. $a_m>2$, stopband and then apply a transformation to the recurrence \labelcref{eq:recRelation} to obtain the analogous connection formulae across an odd, $a_m<-2$, stopband. These results connect \emph{scattering} states $E_m^{\pm}$ across a classically forbidden region and augment the discrete WKB matching procedure used by Braun \cite{Braun1978} to derive the quantization rules for \emph{bound} states.

Since both $E_m^+$ and $E_m^-$ (approximately) satisfy the same recursion relation
\begin{equation}
	\label{eq:recRelationApp}
	E^{\pm}_{m+1}=a_mE^{\pm}_m-E^{\pm}_{m-1}
\end{equation}
near a turning point $m\sim m_{1,2}$, it suffices to consider just $E^+_m$. Approximate discrete WKB expressions for the field $E_m^+$ can be obtained by assuming a solution to the above recurrence relation of the form 
\begin{equation}
	\label{eq:WKBansatz}
	E_m^+=\exp\left\{i\left[ \Phi^0_m + \Phi^1_m + \Phi^2_m + \cdots \right]\right\}
\end{equation}
where the unknown functions $\Phi^j_m$ are labeled according to their rate of change with respect to the index $m$, so that $\dot{\Phi}^j_m=\mathcal{O}\left(m^{-j}\right)$, $\ddot{\Phi}^j_m=\mathcal{O}\left(m^{-j-1}\right)$, and so on. Substituting the discrete WKB ansatz \labelcref{eq:WKBansatz} into the recurrence relation \labelcref{eq:recRelationApp} and solving for the unknown functions $\Phi^j_m$ up to order $\mathcal{O}\left(m^{-1}\right)$ yields \cite{Braun1993, Braun1978}:
\begin{equation}
	\label{eq:WKBeq}
	E_{m}^{+}=\begin{cases}
	\dfrac{1}{\sqrt[4]{4-a_{m}^{2}}}\Bigg[A\exp\left\{ i\int_{m}^{m_{1}}\cos^{-1}\left(\dfrac{a_{k}}{2}\right)\,\mathrm{d}k\right\} +B\exp\left\{ -i\int_{m}^{m_{1}}\cos^{-1}\left(\dfrac{a_{k}}{2}\right)\,\mathrm{d}k\right\} \bigg] & \left(m<m_{1}\right)\\
	\dfrac{1}{\sqrt[4]{a_{m}^{2}-4}}\Bigg[C\exp\left\{ \int_{m_{1}}^{m}\cosh^{-1}\left(\dfrac{a_{k}}{2}\right)\,\mathrm{d}k\right\} +D\exp\left\{ -\int_{m_{1}}^{m}\cosh^{-1}\left(\dfrac{a_{k}}{2}\right)\,\mathrm{d}k\right\} \Bigg] & \left(m_{1}<m<m_{2}\right)\\
	\dfrac{1}{\sqrt[4]{4-a_{m}^{2}}}\Bigg[F\exp\left\{ -i\int_{m_{2}}^{m}\cos^{-1}\left(\dfrac{a_{k}}{2}\right)\,\mathrm{d}k\right\} +G\exp\left\{ i\int_{m_{2}}^{m}\cos^{-1}\left(\dfrac{a_{k}}{2}\right)\,\mathrm{d}k\right\} \Bigg] & \left(m_{2}<m\right)
\end{cases}
\end{equation}

Our goal is to match these solutions across the stopband, i.e.
find a linear relationship between the coefficients $A$, $B$
and $F$, $G$. Performing this matching procedure will require analysis of approximate solutions
to original recurrence \labelcref{eq:recRelation} in the vicinity of each turning turning point, which we obtain by linearizing $a_{m}$ near $m_{1}$ and $m_{2}$. 

\subsection{Even Stopband Connection Formulae}

Suppose we have an $a_{m}>2$ stopband for $m\in\left[m_{1},m_{2}\right]$
(where $m_{1}$, $m_{2}$ are not necessarily integers), surrounded
on either side by propagation bands $\left|a_m\right|<2$.

\subsubsection{Matching Across Even $m=m_{1}$ Turning Point}

When linearized near $m_{1}$, the recurrence \labelcref{eq:recRelation} takes
the form 
\begin{equation}
	E_{m+1}^{+}=\left[2+\alpha_{1}\left(m-m_{1}\right)\right]E_{m}^{+}-E_{m-1}^{+}\quad\left(\text{for \ensuremath{m} near }m_{1}\right)\label{eq:RR1}
\end{equation}
for some $\alpha_{1}>0$ determined by $a_{m}$, or equivalently
\begin{equation}
	E_{m+1}^{+}=\frac{2\nu}{x}E_{m}^{+}-E_{m-1}^{+}\quad\left(\text{for \ensuremath{m} near }m_{1}\right)
\end{equation}
where we have defined $\nu\equiv x+\left(m-m_{1}\right)$ and $x\equiv2/\alpha_{1}$. In
terms of the variables $\nu$ and $x$, the solution to this recurrence
is a linear combination of the Bessel functions $E_{m}^{+}=aJ_{\nu}\left(x\right)+bY_{\nu}\left(x\right)$. Now, the asymptotic expansions of the Bessel functions $J_\nu(x)$ and $Y_\nu(x)$ for large argument $x$ (consistent with small $\alpha_{1}$ and hence slow chirp) and large order $\nu$ (consistent with $m$ far from $m_{1}$) depend on the relative magnitude of $x$ and $\nu$. In each case, the asymptotic expansions are given by (see \cite{Abramowitz},
9.3.2 and 9.3.3): 
\begin{align}
	J_{\nu}\left(x\right) & \approx\begin{cases}
	\sqrt{\dfrac{2}{\pi}}\left(x^{2}-\nu^{2}\right)^{-1/4}\cos\left[-\nu\cos^{-1}\left(\dfrac{\nu}{x}\right)+\sqrt{x^{2}-\nu^{2}}-\dfrac{\pi}{4}\right] & \left(\nu<x\right)\\
	\sqrt{\dfrac{1}{2\pi}}\left(\nu^{2}-x^{2}\right)^{-1/4}\exp\left[-\nu\cosh^{-1}\left(\dfrac{\nu}{x}\right)+\sqrt{\nu^{2}-x^{2}}\right] & \left(\nu>x\right)
	\end{cases}\\
	Y_{\nu}\left(x\right) & \approx\begin{cases}
	\sqrt{\dfrac{2}{\pi}}\left(x^{2}-\nu^{2}\right)^{-1/4}\sin\left[-\nu\cos^{-1}\left(\dfrac{\nu}{x}\right)+\sqrt{x^{2}-\nu^{2}}-\dfrac{\pi}{4}\right] & \left(\nu<x\right)\\
	-\sqrt{\dfrac{2}{\pi}}\left(\nu^{2}-x^{2}\right)^{-1/4}\exp\left[\nu\cosh^{-1}\left(\dfrac{\nu}{x}\right)-\sqrt{\nu^{2}-x^{2}}\right] & \left(\nu>x\right)\end{cases}
\end{align}
We can then express these Bessel function expansions in terms of the original variables $m$, $m_1$, and $\alpha$ by noting $\left(\nu^{2}-x^{2}\right)=\left(a_{m}^{2}-4\right)/\alpha_{1}^{2}$ in the linearized region and applying the change of variable $k\to k+m_{1}-x$ to obtain
\begin{align}
	\nu\cos^{-1}\left(\frac{\nu}{x}\right)-\sqrt{x^{2}-\nu^{2}}=\int_{\nu}^{x}\cos^{-1}\left(\frac{k}{x}\right)\mathrm{d}k & =\int_{m}^{m_{1}}\cos^{-1}\left(\frac{a_{k}}{2}\right)\mathrm{d}k\\
	\nu\cosh^{-1}\left(\frac{\nu}{x}\right)-\sqrt{\nu^{2}-x^{2}}=\int_{x}^{\nu}\cosh^{-1}\left(\frac{k}{x}\right)\mathrm{d}k & =\int_{m_{1}}^{m}\cosh^{-1}\left(\frac{a_{k}}{2}\right)\mathrm{d}k
\end{align}
Combining these results, the Bessel function asymptotic expansions can be rewritten as
\begin{align}
	J_{\nu}\left(x\right) & \approx\begin{cases}
	\sqrt{\dfrac{2\alpha_{1}}{\pi}}\left(4-a_{m}^{2}\right)^{-1/4}\cos\left[-\int_{m}^{m_{1}}\cos^{-1}\left(\dfrac{a_{k}}{2}\right)\mathrm{d}k-\dfrac{\pi}{4}\right] & \left(m<m_{1}\right)\\
	\sqrt{\dfrac{\alpha_{1}}{2\pi}}\left(a_{m}^{2}-4\right)^{-1/4}\exp\left[-\int_{m_{1}}^{m}\cosh^{-1}\left(\dfrac{a_{k}}{2}\right)\mathrm{d}k\right] & \left(m>m_{1}\right)
	\end{cases}\\
	Y_{\nu}\left(x\right) & \approx\begin{cases}
	\sqrt{\dfrac{2\alpha_{1}}{\pi}}\left(4-a_{m}^{2}\right)^{-1/4}\sin\left[-\int_{m}^{m_{1}}\cos^{-1}\left(\dfrac{a_{k}}{2}\right)\mathrm{d}k-\dfrac{\pi}{4}\right] & \left(m<m_{1}\right)\\
	-\sqrt{\dfrac{2\alpha_{1}}{\pi}}\left(a_{m}^{2}-4\right)^{-1/4}\exp\left[\int_{m_{1}}^{m}\cosh^{-1}\left(\dfrac{a_{k}}{2}\right)\mathrm{d}k\right] & \left(m>m_{1}\right)\end{cases}
\end{align}
so that the linear combination of Bessel functions
$E^+_m=aJ_{\nu}\left(x\right)+bY_{\nu}\left(x\right)$
therefore becomes 
\begin{equation}
	\label{eq:m1BesselSol}
	E_{m}^{+}=\begin{cases}
	\dfrac{1}{\sqrt[4]{4-a_{m}^{2}}}\left[\sqrt{\dfrac{\alpha_{1}}{2\pi}}\left(ae^{i\pi/4}-be^{-i\pi/4}\right)\exp\left\{ i\int_{m}^{m_{1}}\cos^{-1}\left(\dfrac{a_{k}}{2}\right)\mathrm{d}k\right\} +\text{C.C.}\right] & \left(m<m_{1}\right)\\
	\dfrac{1}{\sqrt[4]{a_{m}^{2}-4}}\left[-b\sqrt{\dfrac{2\alpha_{1}}{\pi}}\exp\left\{ \int_{m_{1}}^{m}\cosh^{-1}\left(\dfrac{a_{k}}{2}\right)\mathrm{d}k\right\} +a\sqrt{\dfrac{\alpha_{1}}{2\pi}}\exp\left\{ -\int_{m_{1}}^{m}\cosh^{-1}\left(\dfrac{a_{k}}{2}\right)\mathrm{d}k\right\} \right] & \left(m>m_{1}\right)\end{cases}
\end{equation}
as our asymptotic solution for the field $E_{m}^{+}$ near the turning
point $m_{1}$. Comparing this result with the relevant regions of
our discrete WKB solution \labelcref{eq:WKBeq} and eliminating the intermediate
coefficients $a$ and $b$ yields
\begin{align}
	\label{eq:leftMatch}
	A & =  e^{-i\pi/4}\left(\frac{C}{2}+iD\right)\\
	B & =  e^{+i\pi/4}\left(\frac{C}{2}-iD\right)\nonumber
\end{align}
These are the matching relations across the $m=m_1$ turning point.

\subsubsection{Matching Across Even $m=m_{2}$ Turning Point}

We define the even stopband tunneling parameter
\begin{equation} 
\gamma = \int_{m_1}^{m_2}\cosh^{-1}\left(\dfrac{a_{m}}{2}\right)\,\mathrm{d}m.
\end{equation}
Using $\gamma$, we can then rewrite the discrete WKB solution \labelcref{eq:WKBeq} as
\begin{equation}
	E_{m}^{\text{+}}
	=
	\begin{cases}
	\dfrac{1}{\sqrt[4]{4-a_{m}^{2}}}\Bigg[A\exp\left\{ i\int_{m}^{m_{1}}\cos^{-1}\left(\dfrac{a_{k}}{2}\right)\,\mathrm{d}k\right\} +B\exp\left\{ -i\int_{m}^{m_{1}}\cos^{-1}\left(\dfrac{a_{k}}{2}\right)\,\mathrm{d}k\right\} \bigg] & \left(m<m_{1}\right)
	\\
	\dfrac{1}{\sqrt[4]{a_{m}^{2}-4}}\Bigg[De^{-\gamma}\exp\left\{ \int_{m}^{m_{2}}\cosh^{-1}\left(\dfrac{a_{k}}{2}\right)\,\mathrm{d}k\right\} +Ce^{\gamma}\exp\left\{ -\int_{m}^{m_{2}}\cosh^{-1}\left(\dfrac{a_{k}}{2}\right)\;\mathrm{d}k\right\} \Bigg] & \left(m_{1}<m<m_{2}\right)
	\\
	\dfrac{1}{\sqrt[4]{4-a_{m}^{2}}}\Bigg[F\exp\left\{ -i\int_{m_{2}}^{m}\cos^{-1}\left(\dfrac{a_{k}}{2}\right)\,\mathrm{d}k\right\} +G\exp\left\{ i\int_{m_{2}}^{m}\cos^{-1}\left(\dfrac{a_{k}}{2}\right)\,\mathrm{d}k\right\} \Bigg] & \left(m_{2}<m\right)
\end{cases}\label{eq:WKBeq2}
\end{equation}
This reformulation is useful for matching near $m=m_{2}$,
since the discrete WKB solution in the stopband is now expressed as a function
of right turning point $m_{2}$. The recurrence \labelcref{eq:recRelation} linearized near $m=m_{2}$ takes the form 
\begin{equation}
	E_{m+1}^{+}=\left[2-\alpha_{2}\left(m-m_{2}\right)\right]E_{m}^{+}-E_{m-1}^{+}\quad\left(\text{for \ensuremath{m} near }m_{2}\right)
\end{equation}
or equivalently,
\begin{equation}
	\widetilde{E}_{m+1}^{+}=\frac{2\nu}{x}\widetilde{E}_{m}^{+}-\widetilde{E}_{m-1}^{+}
\end{equation}
where we have defined $\nu\equiv x-\left(m-m_{2}\right)$ and $x\equiv 2/\alpha_{2}$. In terms of the variables $\nu$ and $x$, the solution to this recurrence is a
linear combination of the Bessel functions $E_m^+=aJ_{\nu}\left(x\right)+bY_{\nu}\left(x\right)$, just as in the $m=m_1$ case. In particular, the mapping from the $m=m_1$ case is exact under the switch $m\leftrightarrow m_1$ (so as to obtain the modified $m$-dependence of $\nu$) and subsequent replacements $m_1,\alpha_1\to m_2,\alpha_2$. Applying this transformation to the linear combination of asymptotic Bessel functions \labelcref{eq:m1BesselSol} gives
\begin{equation}
	E_{m}^{+}=\begin{cases}
	\dfrac{1}{\sqrt[4]{4-a_{m}^{2}}}\left[\sqrt{\dfrac{\alpha_{2}}{2\pi}}\left(ae^{i\pi/4}-be^{-i\pi/4}\right)\exp\left\{ i\int_{m_{2}}^{m}\cos^{-1}\left(\dfrac{a_{k}}{2}\right)\mathrm{d}k\right\} +\text{C.C.}\right] & \left(m_{2}<m\right)\\
	\dfrac{1}{\sqrt[4]{a_{m}^{2}-4}}\left[-b\sqrt{\dfrac{2\alpha_{2}}{\pi}}\exp\left\{ \int_{m}^{m_{2}}\cosh^{-1}\left(\dfrac{a_{k}}{2}\right)\mathrm{d}k\right\} +a\sqrt{\dfrac{\alpha_{2}}{2\pi}}\exp\left\{ -\int_{m}^{m_{2}}\cosh^{-1}\left(\dfrac{a_{k}}{2}\right)\mathrm{d}k\right\} \right] & \left(m_{2}>m\right)\end{cases}
\end{equation}
as the asymptotic solution for the field $E_{m}^{+}$ near the turning
point $m_{2}$. Comparing this result with the relevant regions of
our discrete WKB solution \labelcref{eq:WKBeq2} and eliminating the intermediate
coefficients $a$ and $b$ yields
\begin{align}
	\label{eq:rightMatch}
	F & = e^{-i\pi/4}\left(Ce^{\gamma}+\frac{iD}{2}e^{-\gamma}\right)\\
	G & = e^{+i\pi/4}\left(Ce^{\gamma}-\frac{iD}{2}e^{-\gamma}\right)\nonumber
\end{align}
These are the matching relations across the $m=m_2$ turning point.

\subsubsection{Matching Across Entire Even Stopband}

Eliminating the coefficients $C$ and $D$ in the matching relations \labelcref{eq:leftMatch,eq:rightMatch} finally gives
\begin{align}
	\label{eq:evenMatch}
	A & = \left(e^{\gamma}+\frac{1}{4}e^{-\gamma}\right)F+i\left(e^{\gamma}-\frac{1}{4}e^{-\gamma}\right)G\\
	B & = -i\left(e^{\gamma}-\frac{1}{4}e^{-\gamma}\right)F+\left(e^{\gamma}+\frac{1}{4}e^{-\gamma}\right)G\nonumber
\end{align}
These are the discrete WKB connection formulae across an entire $a_m>2$ stopband and are identical to the connection formulae across a potential barrier in the standard, continuous WKB approximation \cite{MerzbacherQM}.

\subsection{Odd Stopband Connection Formulae}

Now consider an $a_{m}<-2$ stopband for $m\in\left[m_{1},m_{2}\right]$ (where $m_{1}$, $m_{2}$ are not necessarily integers), surrounded on either side by propagation bands $\left|a_m\right|<2$. We can transform the solution $E^+_m$ in this odd stopband region, given again by \cref{eq:WKBeq}, to a solution in the region near an even stopband by taking
\begin{align}
a_{m} & \to \widetilde{a}_{m}\equiv-a_{m}\\
E_{m}^{+} & \to \widetilde{E}_{m}^{+}\equiv\left(-1\right)^{m}E_{m}^{+}\nonumber
\end{align}
which leaves the recursion relation \labelcref{eq:recRelation} unchanged. This transformation can be achieved by multiplying \labelcref{eq:WKBeq} by $(-1)^m$ and using the identities $\cos^{-1}\left(a_m/2\right)=\pi-\cos^{-1}\left(\widetilde{a}_m/2\right)$ and $\left(-1\right)^{m}e^{\pm im\pi}=1$. Effecting this transformation yields
\begin{equation}
	\label{eq:tildeWKBeq}
	\widetilde{E}_{m}^{+}=\begin{cases}
	\dfrac{1}{\sqrt[4]{4-\widetilde{a}_{m}^{2}}}\Bigg[\widetilde{A}\exp\left\{ i\int_{m}^{m_{1}}\cos^{-1}\left(\dfrac{\widetilde{a}_{k}}{2}\right)\,\mathrm{d}k\right\} +\widetilde{B}\exp\left\{ -i\int_{m}^{m_{1}}\cos^{-1}\left(\dfrac{\widetilde{a}_{k}}{2}\right)\,\mathrm{d}k\right\} \bigg] & \left(m<m_{1}\right)\\
	\dfrac{1}{\sqrt[4]{4-\widetilde{a}_{m}^{2}}}\Bigg[\widetilde{F}\exp\left\{ -i\int_{m_{2}}^{m}\cos^{-1}\left(\dfrac{\widetilde{a}_{k}}{2}\right)\,\mathrm{d}k\right\} +\widetilde{G}\exp\left\{ i\int_{m_{2}}^{m}\cos^{-1}\left(\dfrac{\widetilde{a}_{k}}{2}\right)\,\mathrm{d}k\right\} \Bigg] & \left(m_{2}<m\right)\end{cases}
\end{equation}
where we have defined
\begin{align}
	\label{eq:tildeCoeffs}
	\widetilde{A} & = e^{-im_{1}\pi}B\\
	\widetilde{B} & = e^{+im_{1}\pi}A\nonumber\\
	\widetilde{F} & = e^{-im_{2}\pi}G\nonumber\\
	\widetilde{G} & = e^{+im_{2}\pi}F\nonumber
\end{align}
Since the transformed variable $\widetilde{a}_m$ describes an even $\widetilde{a}_m\equiv-a_m>2$ stopband, we can apply the even stopband connection formulae \labelcref{eq:evenMatch} to the transformed discrete WKB solution \labelcref{eq:tildeWKBeq} and match the transformed coefficients $\widetilde{A},\ldots,\widetilde{G}$. Applying this even stopband matching and subsequently transforming back to the original coefficients $A,\ldots, G$ finally gives
\begin{align}
	\label{eq:appMatching}
	B & = e^{+im_{1}\pi}\left[e^{-im_{2}\pi}\left(e^{\gamma}+\frac{1}{4}e^{-\gamma}\right)G+ie^{+im_{2}\pi}\left(e^{\gamma}-\frac{1}{4}e^{-\gamma}\right)F\right]\\
	A & = e^{-im_{1}\pi}\left[-ie^{-im_{2}\pi}\left(e^{\gamma}-\frac{1}{4}e^{-\gamma}\right)G+e^{+im_{2}\pi}\left(e^{\gamma}+\frac{1}{4}e^{-\gamma}\right)F\right]\nonumber
\end{align}
where
\begin{equation}
	\gamma = \int_{m_1}^{m_2}\cosh^{-1}\left(\dfrac{\widetilde{a}_{m}}{2}\right)\,\mathrm{d}m = \int_{m_1}^{m_2}\cosh^{-1}\left(-\dfrac{a_{m}}{2}\right)\,\mathrm{d}m.
\end{equation}

These are the discrete WKB connection formulae across an entire odd stopband and are precisely the relations contained in \cref{eq:matching}. Interestingly, odd stopbands have no analogy in the continuous WKB approximation, and the odd stopband connection formulae \labelcref{eq:matching} are not identical to the even stopband/continuous WKB connection formulae \labelcref{eq:evenMatch}. In particular, the presence of additional turning point phases $\exp\left(\pm im_{1,2}\pi\right)$ in the odd stopband connection formulae \labelcref{eq:appMatching} has important consequences in quantum mechanical applications \cite{Braun1993}.

\pagebreak
\twocolumngrid
\section{TEM Image Analysis}
\label{sec:imaging}

\begin{figure}[b!]
	\centering
	\includegraphics[width=\columnwidth]{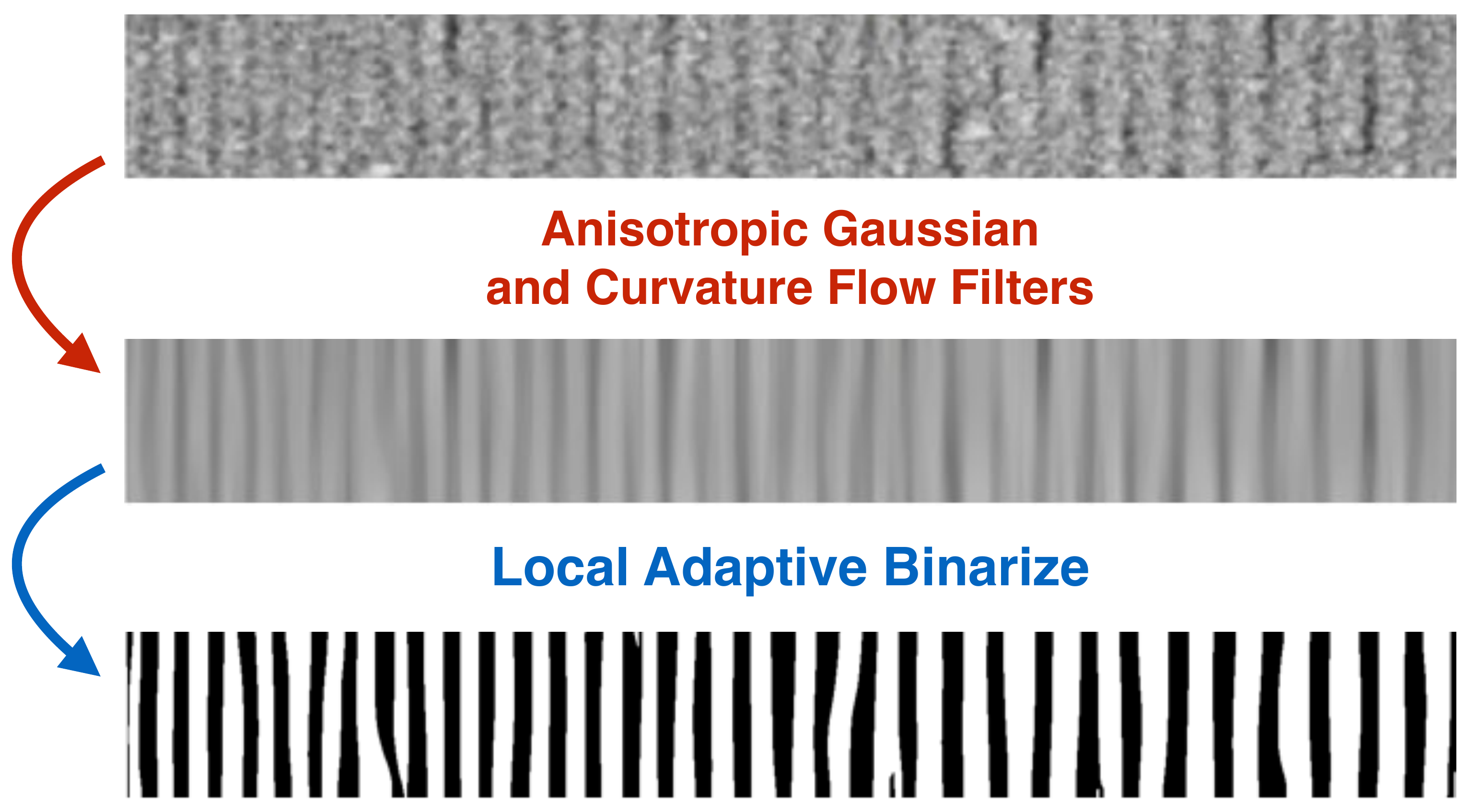}
	\caption{
	Sketch of our TEM image analysis procedure applied to the ACPC in the elytra of the golden \emph{C. aurigans} beetle. Original TEM image (top) taken from Ref. \cite{Libby2014}. }
	\label{fig:imagingEx}
\end{figure}

In the main text, we compare the bilayer number $N$ found in several ACPCs in nature to our predicted minimal number of bilayers $N^{\gamma\sim1}_\text{exp/lin}$ in the case of exponential and linear chirp. This comparison requires an estimate of the optical thickness ratio 
\begin{equation}
\label{eq:Xi}
	\Xi=\left(\ell_{L}\right)_{m}/\left(\ell_{H}\right)_{m}=n_{L}\left(d_{L}\right)_{m}/n_{H}\left(d_{H}\right)_{m},
\end{equation} 
which we estimate for each biological ACPC via analysis of TEM images given in the relevant reference. In this section we detail the image analysis process, a sketch of which is shown in \cref{fig:imagingEx}, using the elytra of the golden \emph{Chrysina aurigans} beetle \cite{Libby2014} as an example.

First, an anisotropic Gaussian filter is applied to the TEM image along the assumed axis of the ACPC, which smooths the image while highlighting the low- and high-index layers. Then a curvature flow filter is applied in order further smooth the image while preserving the edges between low- and high-index layers. Finally we binarize the image, where the binarization thresholds are determined  locally in the image in order isolate the the low- and high-index layers. 

\begin{figure}[b!]
	\centering
	\includegraphics[width=\columnwidth]{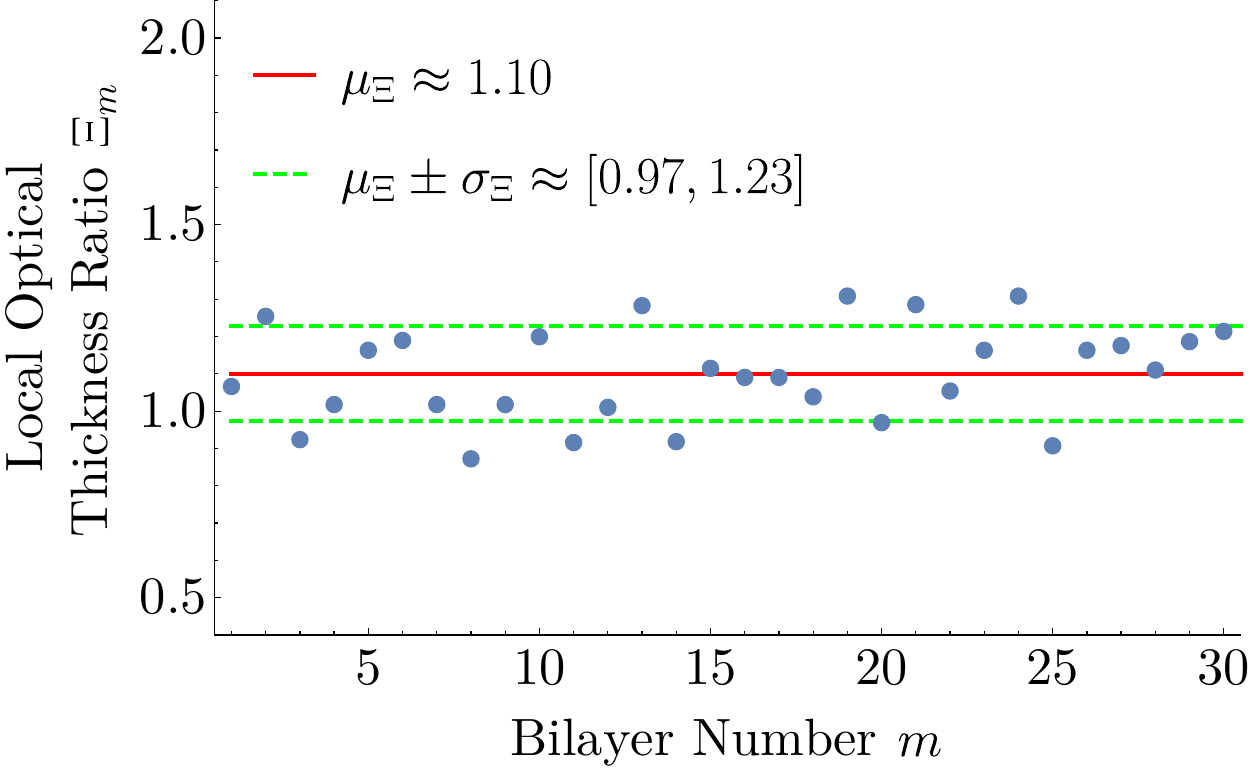}
	\caption{
	Plot of the local optical thickness ratios $\Xi_m$ calculated for each bilayer $m$ in the \emph{C. aurigans} TEM image shown in \cref{fig:imagingEx}. The $\Xi_m$ values for the first and last measured bilayers are discared to account for possible incomplete cropping of the original TEM image. The mean optical thickness ratio $\mu_\Xi$ and the standard deviations $\mu_\Xi\pm\sigma_\Xi$ about the mean are also shown.}
	\label{fig:imagingEx2}
\end{figure}

After binarization, we take a thin (few pixel high) horizontal cross-section of the image, from which the layer thicknesses $\left(d_{L/H}\right)_m$ and hence local optical thickness ratios $\Xi_m$ \labelcref{eq:Xi} can be determined. The value of $\Xi_m$ for each bilayer $m$ in the \emph{C. aurigans} TEM image is plotted in \cref{fig:imagingEx2}. The fixed ratio $\Xi$ is finally estimated by taking the average optical thickness ratio $\Xi=\mu_\Xi$ over all the ACPC bilayers $m$.

% Bibliography
%\bibliography{CookAmir2016.bib}

\end{document}